\documentclass[journal,10pt]{IEEEtran}
\usepackage[T1]{fontenc}
\usepackage{cite}
\usepackage{enumitem}
\usepackage{mdwlist}
\usepackage{url}
\usepackage{amssymb}
\usepackage{amsfonts}
\usepackage{fancyhdr}  %
\usepackage{extarrows}
\usepackage{algorithm}
\usepackage{multirow,tabularx}
\usepackage{mathtools}
\usepackage{xcolor}
\usepackage[english]{babel}
\usepackage{caption}
\usepackage{bm}
\usepackage{algorithmic}
\usepackage{amsmath}
\usepackage{subfigure}
\usepackage{makecell}
\usepackage{colortbl}
\usepackage{booktabs}

\usepackage{hyperref} 
\usepackage{cleveref} 
\usepackage{setspace}
\hypersetup{colorlinks = true, 
	    linkcolor = black, 
	    urlcolor = black,
            citecolor = black} 

\ifCLASSINFOpdf
  \usepackage[pdftex]{graphicx}
  \graphicspath{{../pdf/}{../jpeg/}}
  \DeclareGraphicsExtensions{.pdf,.jpeg,.png}
\else
  \usepackage[dvips]{graphicx}
  \graphicspath{{../eps/}}
  \DeclareGraphicsExtensions{.eps}
\fi
\begin{document}
\title{Big AI Models for 6G Wireless Networks: Opportunities, Challenges, and Research Directions}

\author{Zirui~Chen,
		Zhaoyang~Zhang,
		and Zhaohui~Yang
  \thanks{This work was supported in part by National Natural Science Foundation of China under Grants 62394292 and U20A20158, Ministry of Industry and Information Technology under Grant TC220H07E, Zhejiang Provincial Key R\&D Program under Grant 2023C01021, and the Fundamental Research Funds for the Central Universities. (\textit{Corresponding Author: Zhaoyang~Zhang})}
  \thanks{Z.~Chen, Z.~Zhang and Z.~Yang are with College of Information Science and Electronic Engineering, Zhejiang University, Hangzhou 310027, China, and with Zhejiang Provincial Key Laboratory of Info. Proc., Commun. \& Netw. (IPCAN), Hangzhou 310007, China. Z. Yang is also with Zhejiang Lab, Hangzhou, 311121, China. (e-mail: \{ziruichen, ning\_ming, yang\_zhaohui\}@zju.edu.cn)}
}

\maketitle
\vspace{-0.3em}
\begin{abstract}
Recently, big artificial intelligence  models (BAIMs) represented by chatGPT have brought an incredible revolution. With the pre-trained BAIMs in certain fields, numerous downstream tasks can be accomplished with only few-shot or even zero-shot learning and exhibit state-of-the-art performances. As widely envisioned, the big AI models are to rapidly penetrate into major intelligent services and applications, and are able to run at low unit cost and high flexibility. In 6G wireless networks, to fully enable intelligent communication, sensing and computing, apart from providing other intelligent wireless services and applications, it is of vital importance to design and deploy certain wireless BAIMs (wBAIMs). However, there still lacks investigation on architecture design and system evaluation for wBAIM. In this paper, we provide a comprehensive discussion as well as some in-depth prospects on the demand, design and deployment aspects of the wBAIM. We opine that wBAIM will be a recipe of the 6G wireless networks to build high-efficient, sustainable, versatile, and extensible wireless intelligence for numerous promising visions. Then, we provide the core characteristics, principles, and pilot studies to guide the design of wBAIMs, and discuss the key aspects of developing wBAIMs through identifying the differences between the existing BAIMs and the emerging wBAIMs. Finally, related research directions and potential solutions are outlined.

\end{abstract}

\IEEEpeerreviewmaketitle

\section{Introduction} \label{introduction}
\begin{figure*}
  \centering
  \includegraphics[width=0.95\linewidth]{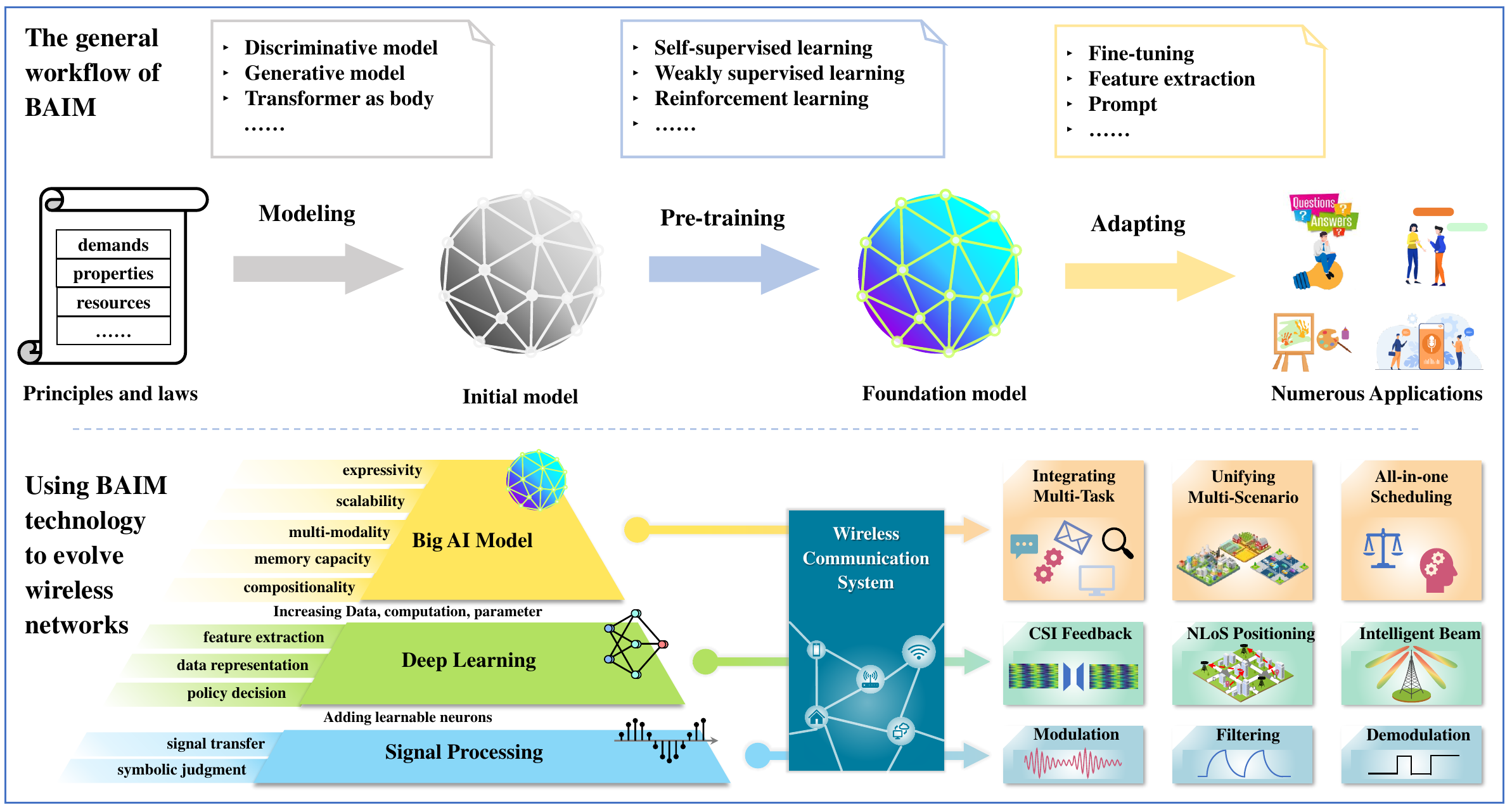}
  \vspace{0.3em}
     \caption{\small { 
     BAIM, also termed as foundation model, is a new machine learning paradigm that captures broad-adaptable intelligence and provides numerous downstream applications through appropriate modeling, sufficient pre-training and customized adaptations.
     Establishing BAIM dedicated to wireless networks will enable an evolution of wirless systems toward more integrated functions, more flexible architectures, and more differentiated services. }
     }
  \label{fig_overview}
\end{figure*}

As an emerging machine learning paradigm, big artificial intelligence model (BAIM), also termed as foundation model, is revolutionizing various fields, including natural language processing (NLP). Leveraging powerful model, huge parameter scale, abundant data, and massive computational resources, BAIMs gain unprecedented generalization and broad-adaptable intelligence from pre-training. Just one pre-trained BAIM can be adapted to a wide range of downstream applications and achieve state-of-the-art performance by fine-tuning, few-shot, or even zero-shot learning. In summary, Fig. \ref{fig_overview} presents the general workflow and popular algorithms of BAIM technology. 
This amazing capability of BAIM overcomes  the limitations of past deep neural network (DNN), which is weakly generalizable and  can only achieve task-specific intelligence. 
DNN now can not only be used as an alternative for specific  functional modules, but, especially BAIM, has the potential to bring universal information processing intelligence in at least one whole field. 

Advances in information processing technology will profoundly empower wireless network's evolution, which can be broadly categorized into three levels, as shown in Fig. \ref{fig_overview}.
In the first level, effective tradition signal processing techniques realizes modulation, filtering, and demodulation, guaranteeing basic wireless transmission.
Then, in the second level, deep learning-assisted wireless communication achieves implicit feature extraction and high-dimensional representation, arising low-cost channel feedback, non line-of-sight (NLoS) positioning, and intelligent beam.
These developments facilitate advanced wireless applications such as multiple-input multiple-output (MIMO) transmission and reliable wireless localization.
Further, if a BAIM can be generated for wireless networks and fruit broad-adaptable intelligent wireless information processing, it will significantly contribute to the promising visions in 6G: building ubiquitous (integrating multi-task, unifying multi-scenario, and all-in-one scheduling) intelligence to support unprecedented usage cases \cite{vision_6G}.

The recent surge in machine learning and the significance of integrating advanced information processing technologies and wireless systems motivate researching BAIM for 6G networks. 
There are two evidences for the practicability of such research. First, existing researches on wireless AI \cite{csi_positioning,csinet,reinforcement_wireless} have shown that the advantages of DNNs in nonlinear transformation, robustness, and decision-making can be well utilized to solve problems for wireless networks, and thus higher intelligence is likely to bring more opportunities to wireless networks. Second, the massive and informative wireless data, the rich computational resources from ubiquitous wireless devices, and the continuous algorithm accumulation of wireless AI also provide critical support to generate BAIMs belonging to wireless networks.

In view of the above expectations, this paper provides a prospect on the wireless BAIM (wBAIM) for 6G networks. Different from \cite{AIGC_network}, which surveys the applications of existing language and vision BAIMs in mobile networks, this paper investigates BAIMs dedicated to wireless networks. This paper first introduces the opportunities, design and some pilot studies of wBAIMs to meet the vital demands of 6G networks. Then, this paper discusses the open challenges and key problems of wBAIMs. Moreover, some research directions and potential solutions are present. Finally, the conclusion is drawn.

 \section{Why the Wireless BAIM is Needed in 6G?}
\subsection{What Kind of Wireless Intelligence is Indispensable in 6G?}

\begin{figure*}
  \centering
  \includegraphics[width=0.85\linewidth]{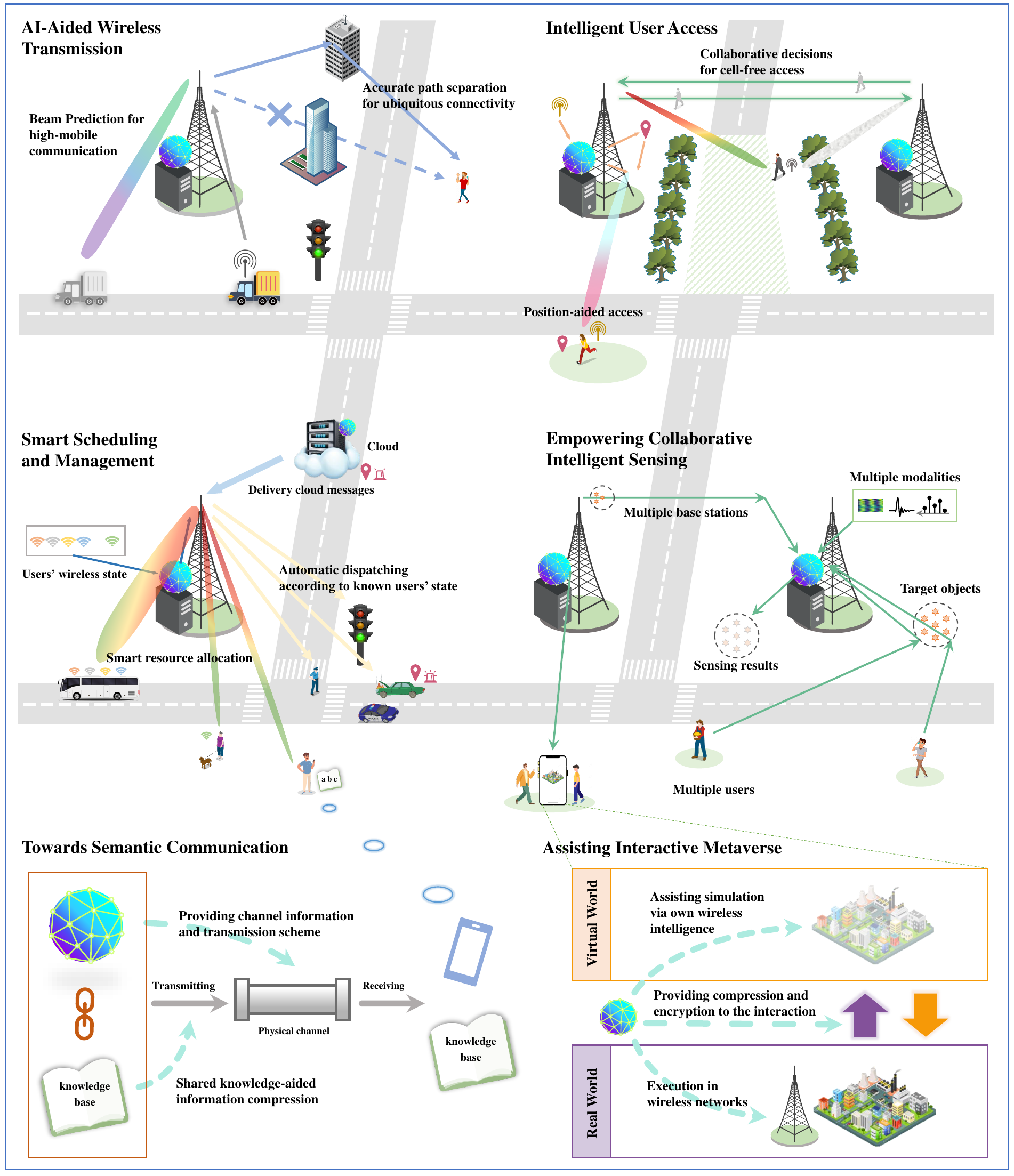}
  \vspace{0.3em}
      \caption{\small The numerous use scenarios in 6G wireless networks place urgent demands for high-efficient, sustainable, versatile, and extensible wireless intelligence. This required intelligence is envisioned to be built through the wBAIM.}
  \label{fig_application}
\end{figure*}

International telecommunication union (ITU) identifies six typical use scenarios of 6G: immersive communication, artificial intelligence (AI) and communication, hyper-reliable and low latency communication, ubiquitous connectivity, massive communication and integrated sensing and communication. To realize these promising use scenarios, 6G requires stronger technical means to provide richer functional support, and one commonly applicable design principle is to build ubiquitous intelligence in wireless networks \cite{vision_6G}.
Next, we illustrate what kind of wireless intelligence is exactly indispensable through some required functions in 6G, as summarized in Fig. \ref{fig_application}.

\subsubsection{AI-Aided Wireless Transmission}
AI-Aided transmission is a vital goal of AI and communication and also plays an important role in ubiquitous connectivity, which requires accurate analysis of the user's wireless state and adaptively setting up transceiver schemes. For example, to enable reliable transmission at high mobility, the expectant wireless AI should be able to infer motion status and perform accuracy channel prediction. Besides, for wide coverage in complex environments, wireless intelligence requires accurate multi-path separation and beamforming based on the received reference signal.

\subsubsection{Intelligent User Access} 
Wireless intelligence is significant for efficient user access to realize hyper reliable and low-latency communication. Intelligent user access is inseparable from making adaptive decisions, synchronizing information between multiple scenarios, and mapping/combining wireless signals with nature space. For example, by inferring the user position from wireless signals, a beam with high-precision spatial directivity can be directly established rather than through randomly scan, enabling faster and less resource-intensive access. In addition, for cell-free access, neighboring cells should synchronize the user motion status and collaboratively make decisions to select appropriate BS for service.

\subsubsection{Smart Scheduling and Management} 
The future of scheduling and management is bound to be intelligent, which will significantly facilitate massive communication. Cooperated information processing and decision-making are necessary for this function. Leveraging intelligence and real-time state information, wireless system can establish resource allocation schemes that exactly balance efficiency and fairness, such as increasing transmission power for areas with high user density and assigning suitable sub-bands to users for transmission. In addition, future city management, such as automatic dispatching when dealing with emergencies, also requires cloud-edge message synchronization and intelligent decision-making.

\subsubsection{Empowering Collaborative Intelligent Sensing} 
The empowerment from AI technology is important in achieving integrated sensing and communication. Due to the limited information in single wireless signal, collaboratively processing of multi-modal, multi-user, and multi-scenario signals is a critical guarantee of high-quality sensing. Specifically, collaborative learning of multi-modal signals facilitates multi-dimensional and multi-granularity sensing; combining observations from users at different positions will improve the sensing accuracy; information interaction between multiple base stations (BS) can help filter environmental interference to achieve large-scale target reconstruction.

\subsubsection{Assisting Interactive Metaverse} 
Metaverse is a promising technical means for immersive communication, which requires comprehensive supports for the simulation in virtual world, the mapping in real world, and the information interaction between virtual and real world. Specifically, an intelligent machine that can simulate wireless states in virtual world is necessary. Meanwhile, the capacities of autonomously making decisions and adjustments according to simulation results is also required. Besides, it is also valuable to provide high-efficient transmission for the interaction between virtual and real world, including information compression and encryption.

\subsubsection{Towards Semantic Communication} 
Semantic communication can further enhance communication efficiency and provide strong support for massive communication. It integrates user requirements and known information into the communication process, which often needs the help of AI technology. Analyzing user's wireless state and accordingly formulates the transmission or decoding methods to constitute semantic-level communication at the physical layer is an essential function. In addition, to couple with diverse knowledge bases, adequate information capacity and intelligence level is also necessary for wireless model.

In summary, to support numerous promising visions \cite{vision_6G}, the wireless AI necessary for 6G needs to efficiently extract, transform, and represent the wireless signal, make adaptive decisions, and map/combine the wireless signal with information in natural space, while being able to interact and collaborate information from multiple modalities, multiple users, and multiple scenarios, and automatically combine multiple functions for advanced applications.

\subsection{Why Using BAIM as a Recipe of 6G?}
There have been some preliminary wireless AI technologies with performance gains on corresponding use cases. However, the current architecture of wireless AI is still task-specific and scenario-specific, requiring corresponding data collection and training according to the applied task and considered scenario, as shown in Fig. \ref{fig_architecture}. These characteristics prevent existing architectures from remaining high-efficiency in complex use cases and autonomously integrating functions of multiple tasks and scenarios to provide advanced applications. In addition, paying extensive data and training costs for every task and scenario makes the widespread deployment of existing wireless AI unsustainable. Moreover, there are often resource conflicts and interaction barriers between current wireless AI models since they all focus on their specific tasks and scenarios. Therefore, the existing wireless AI architecture still cannot meet the required wireless intelligence in 6G networks. Establishing high-efficient, sustainable, versatile, and extensible wireless intelligence is imperative.

“High-efficient, sustainable, versatile and extensible” are exactly the features that significantly distinguish BAIMs from past AI technology, making the BAIM likely to be the key recipe of the next-generation wireless networks. Building a BAIM capable of characterizing the intrinsic hidden features of extremely complex wireless channels, environments, and objects, and tightly integrating it into wireless networks, will present numerous exciting opportunities and satisfy the requirements of the above use cases, as shown in Fig. \ref{fig_application}. Next, we will introduce how the wBAIM reshapes future wireless systems in Section \ref{section_architecture}.

\section{How wBAIM Works in 6G?} \label{section_architecture}

wBAIM-based architecture pursues better versatility and extensibility than existing wireless architectures, and establishing a unified deployment paradigm with the support of pre-training is its core characteristic.

\subsection{Pre-Training a wBAIM as a Foundation Model}\label{pretraining}
\begin{figure*}
  \centering
  \includegraphics[width=0.95\linewidth]{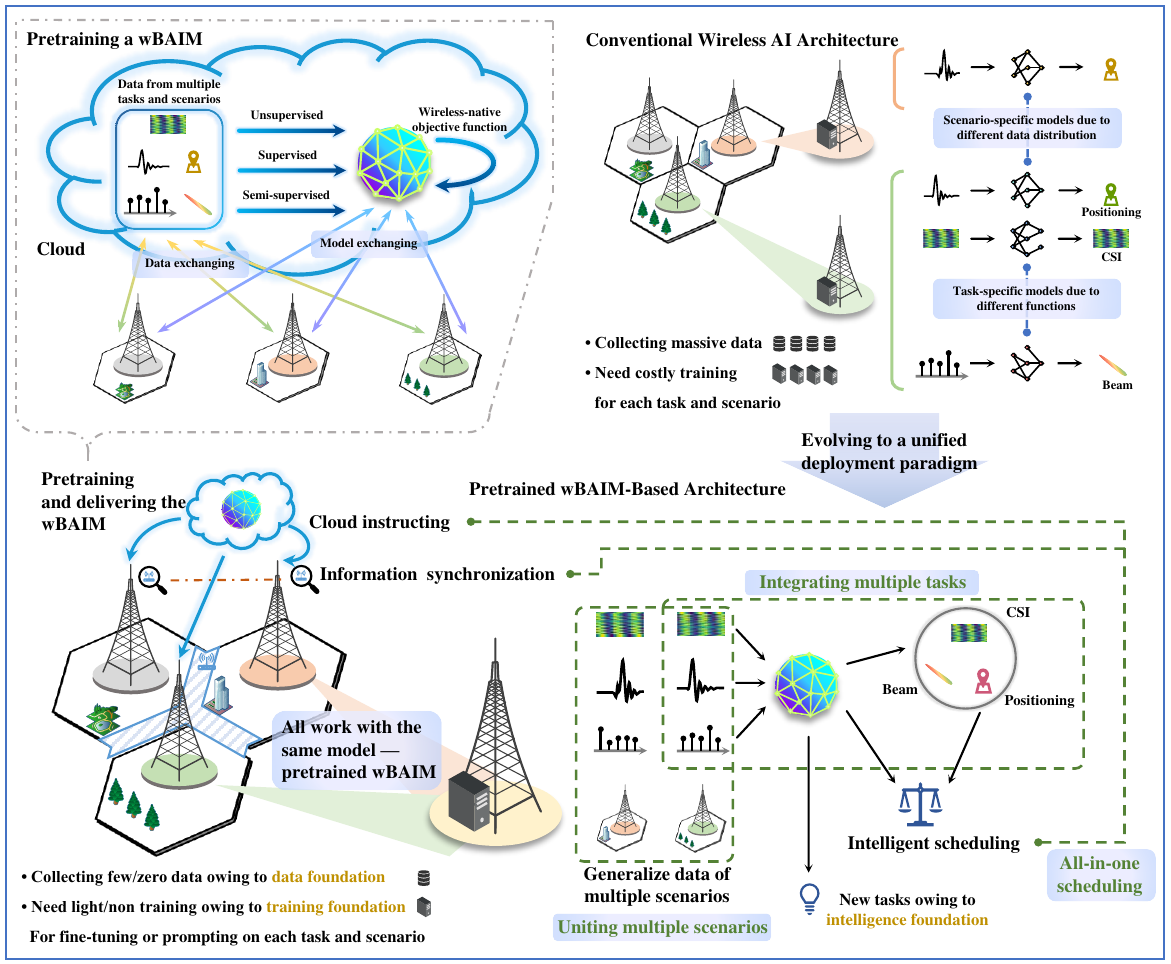}
  \vspace{0.3em}
      \caption{\small The distinguishing features of wBAIM-based architecture are as following. First pre-training a wBAIM that captures universal wireless intelligence, and then providing a unified paradigm in deployment based on the wBAIM: integrating multiple wireless tasks, unifying multiple communication scenarios, and network-wide all-in-one scheduling.}
  \label{fig_architecture}
\end{figure*}

Pre-training is an important feature of BAIMs, which is also of the wBAIM. Rather than performing task- and scenario-specific training on targeted deployed devices, the wBAIM is pre-trained, e.g., in the collaboration of cloud and edge, as shown in Fig. \ref{fig_architecture}. The primary aim of wBAIM's pre-training is to generate a model that only requires fine-tuning or prompting when deployed to numerous downstream wireless tasks and scenarios. This kind of broad-adaptable intelligence has been initially implemented by BAIMs in NLP, and it is also the fundamental essence of wBAIM. To achieve this, data from multiple wireless scenarios and tasks, training methods adapted to wireless data types such as unsupervised learning for channel state information (CSI) data without labels and supervised learning for labeled channel-position pairs, and wireless-native objective functions are indispensable. The advantages of pre-training are twofold. Firstly, with more wireless data and richer training resources, the intelligence of the wireless model will be significantly improved, ensuring ultra performance in vast use cases. Secondly, based on the pre-trained model, extensive data collection and training on specific tasks and scenarios will no longer be required, reducing the total overhead of widespread deployment.

\subsection{A Unified Paradigm in AI Deployment}
Focusing resources on pre-training the wBAIM is not just aimed at improving performance. Its most significant significance lies in breaking the barriers between tasks, scenarios, and scheduling, as illustrated in Figure \ref{fig_architecture}. By bridging tasks, scenarios, and scheduling, the wireless system becomes more versatile in functionality and enhances its expansibility for new use cases.

\subsubsection{Integrating Multiple Wireless Tasks}
Integrating multiple wireless tasks is reasonable and promising, as these tasks are highly interrelated. For instance, CSI feedback requires feature extraction and characterization of channel data, while channel-based positioning also necessitates analogous feature extraction. Besides, the intelligent beam also relies on channel characterization. 
The beauty is that achieving multiple related tasks based on only one model through low-cost fine-tuning or prompting is exactly one of the distinguishing features of BAIMs.
Thus, wBAIM-based architecture emphasizes integrating multiple regular wireless tasks based on the pre-trained wBAIM rather than operating each task in isolation on a specific model. For example, the processing of noisy reception pilot, compressed channel and signal size, and the inference of user location are all integrated. This versatility in multiple basic functions will also facilitate numerous advanced applications.

\subsubsection{Unifying Multiple Communication Scenarios}
If different scenarios are mostly deployed with differentiated models, the cost of extending new scenarios, the information interaction between scenarios, and the dynamic changes of scenarios are all very tricky. Thus, using one model to unify multiple served scenarios is significant.
Although different scatterer locations in different scenarios result in different data distributions, the electromagnetic wave consistently adheres to the same physical laws. These constant laws guarantee the feasibility of using one neural network to serve multiple scenarios, but the learning network requires a big learning capacity. This corresponds to another feature of wBAIM-based architecture: serving all scenarios via the big pre-trained model with the cross-scenario generalization instead of providing different scenarios with different models. Specifically, the cross-scenario generalization seeks to unify scenarios with different scatterer locations on the spatial scale, as well as the dynamically changing scenarios on the time scale. 

\subsubsection{Network-Wide All-in-One Scheduling}
Integrating tasks and unifying scenarios results in a network-wide all-in-one scheduling that includes intra-cellular autonomy, cross-scenario synchronization, and cloud-based instruction. Specifically, by integrating multiple tasks, the system can acquire multi-modal state information from multiple users within a cell, allowing it to provide real-time resource allocation. Additionally, since different scenarios deploy the same model, the information obtained by different scenarios' models is consistent in format. Therefore, it is easy to synchronize information between scenarios, which brings convenience for intelligent scheduling across scenarios. Similarly, the cloud that pre-trains the wBAIM can also easily exchange information with the scenarios' wBAIM. As a result, the cloud instructions are also easily understood and executed by the model deployed in each scenario. This all-in-one architecture effectively avoids resource conflicts between multiple intelligent tasks and enables seamless service switching across scenarios.

\subsection{Pilot Studies of Pre-Training and Unified Deployments}

\begin{figure*}
  \centering
  \includegraphics[width=0.95\linewidth]{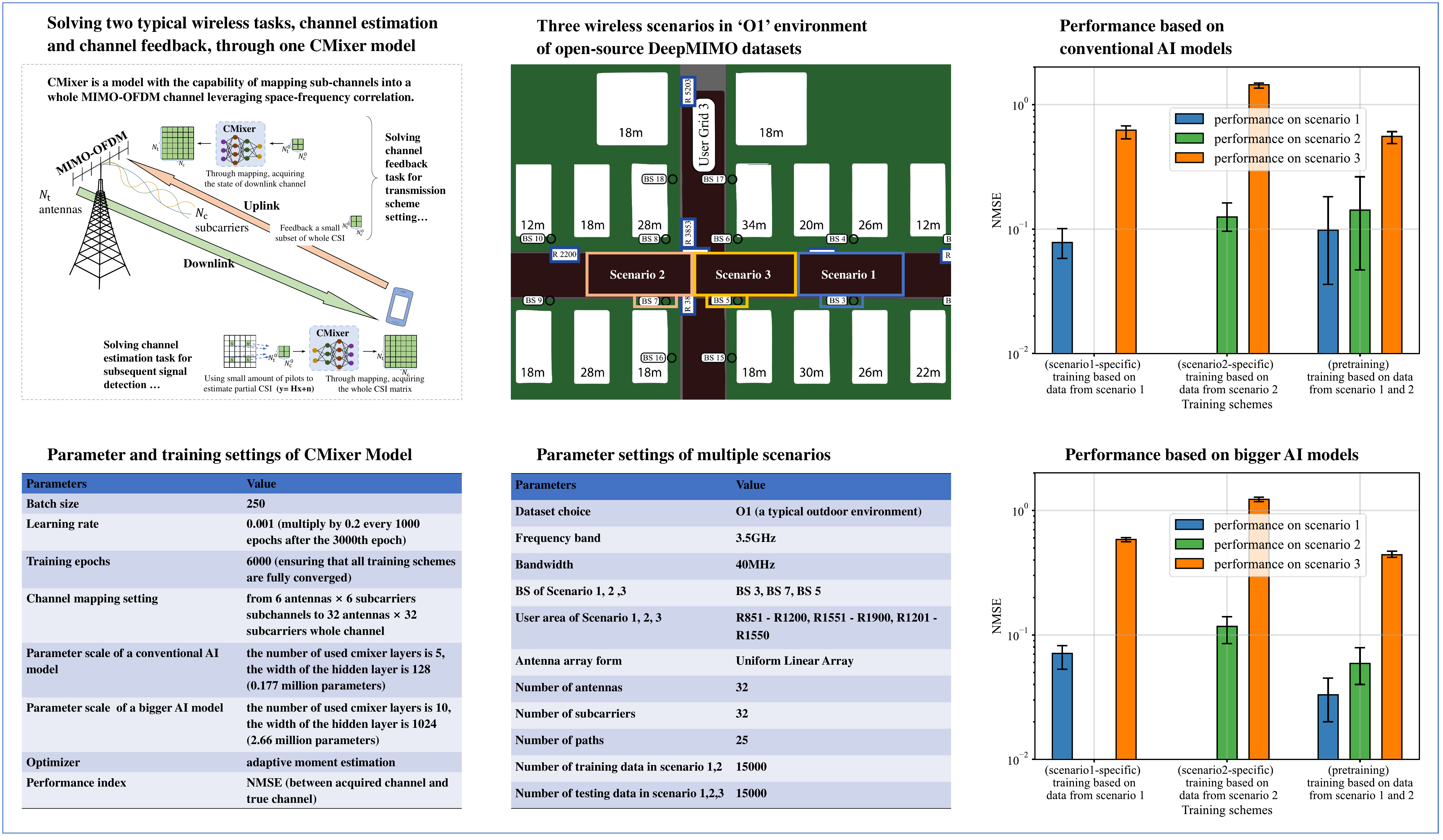}
      \caption{\small A graphic illustration of pilot studies in wireless pre-training and unified deployments. The left part shows how to use a CMixer \cite{cmixer} model to serve both channel estimation and feedback tasks and provides the parameter settings of this model. The middle part shows the communication scenarios and system settings from the open source dataset DeepMIMO \cite{deepmimo}. We assume that the training data is collected from scenario 1 and 2 and the trained models are deployed in scenario 1, 2 and 3. The right part shows the test performance in each scenario with 10 independent repeated experiments. The colorful bars are the mean values and the error bars are upper and lower limits.}
  \label{fig_experiment}
\end{figure*}

This subsection takes a demo experiment to preliminarily demonstrate the feasibility and value of pre-training to provide a unified deployment. Specifically, we model a complex-domain multilayer perceptron (MLP)-Mixer (CMixer) \cite{cmixer} neural network, which can map the whole MIMO-OFDM channel based on some sub-channels. Meanwhile, we use data collected from two scenarios to train this model. 
Fig.~\ref{fig_experiment} details this demo experiment, including transceiver architecture, scenario graph, and parameter tables. Overall, this a simplified pre-training, but can still yield more unified deployments than task- and scenario-specific training. First, since channel mapping needs the model to deeply mine the correlation in space and frequency domains, an effective mapping model can be used for two typical tasks, low-pilot channel estimation and low-cost channel feedback \cite{cmixer}. In addition, training data from multiple scenarios can provide richer learnable knowledge than that from a single scenario. And with the adequate learning capacity, more extensive learning often improves performance and generalization.

Numerical results are shown in the bar graphs of Fig.~\ref{fig_experiment}. The CMixer model achieves high-accuracy channel mapping, effectively assisting the transceivers. Only using a small number of pilots to estimate some sub-channels, and inputting these sub-channels into CMixer can output the whole channel. As well, 
with only partial feedback subchannels,
the BS can reconstruct the whole downlink channel with low normalized mean square error (NMSE) based on the CMixer model. This eliminates the need to design separate models for these two tasks, as well as the corresponding data and training overheads. In addition, the comparisons between scenario-specific training and pre-training show that a single model can be successfully generalized to multiple scenarios. Especially, with the support of sufficient parameter scales (bigger model), pre-training can achieve better performance compared to scenario-specific training, both in scenarios that collected the training data (scenarios 1 and 2) and in new scenario (scenario 3). These results corroborate our perspectives, demonstrating that the wBAIM based on pre-training can support unified deployments across tasks and scenarios, and provide better performance than conventional wireless AI schemes.

\section{Challenges and Key Problems Relating wBAIMs}\label{section_challenges}
\begin{figure}
  \centering
  \includegraphics[width=0.98\linewidth]{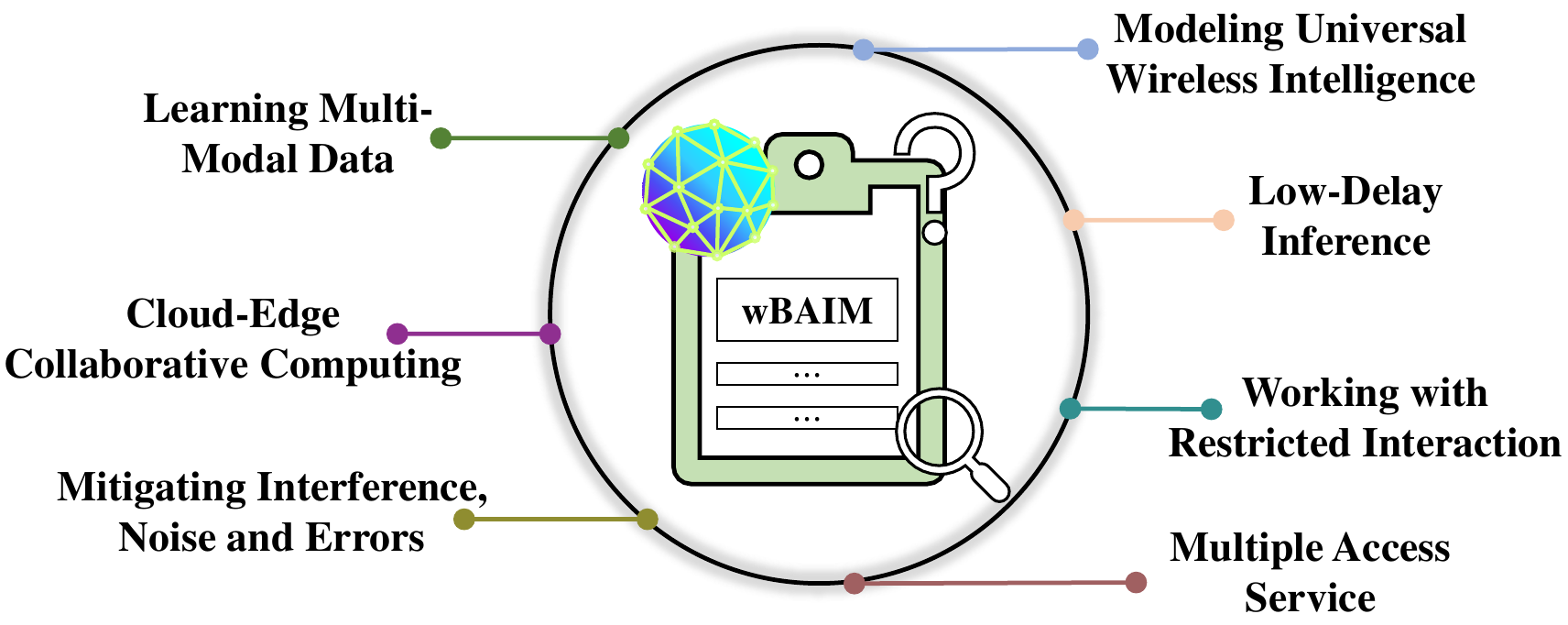}
      \caption{\small The challenges and key problems of the wBAIM.}
  \label{fig_challenge}
\end{figure}

Despite some initial successful BAIMs in other fields and the above potential pilot studies, wBAIMs still face some wireless-constraint challenges and key issues due to the unique properties and requirements of wireless networks, as summarized in Fig. \ref{fig_challenge}.

\subsection{Modeling and Capturing Universal Wireless Intelligence}
The primary challenge is the intelligent form that can serve the whole system, including numerous distinguished wireless tasks and scenarios. Generally, the form must drives to learn the fundamental mechanism that generates various regular phenomena and satisfies numerous application requirements. For instance, GPT \cite{gpt3} uses continuous context generation to mine the core intelligence of language, close correlation between contexts, and meets the major requirements of language, conversation. In wireless networks, the regulations of various tasks and scenarios are essentially derived from the physical laws of electromagnetic waves, and the core requirements all lie in acquiring wireless states and adaptive decision-making. These principles will guide the modeling of universal wireless intelligence. However, the specific form and learning paradigms are still the open and key problems of wBAIM research.

\subsection{Learning and Representing Multi-Modal Data}
The types of wireless data are diverse, which needs the wBAIM to cope with the input and output in various modalities, such as channel frequency response (CFR), position coordinates, and received signal. Different modalities have different data structures and characteristics, challenging the function and generalization of the wireless model. In addition, there are unique correlations and complementarities between different wireless modalities. How to exploit these properties, such as fusing state information from CFR for more accurate received signal detection and complementing spatial information from position coordinates for generating high-quality beam vectors, is also an important problem.

\subsection{Cloud-Edge Collaborative Computing}
The rich data and computing resources in wireless networks lay the foundation for generating the wBAIM. However, how to fully utilize these resources still needs studies. On one hand, the problem of data silos needs to be solved. Not the data in all scenarios can be centralized in the cloud, limited by transmission costs, privacy, and other issues. Effective information exchange methods are necessary to enable the edge data to participate in training. On the other hand, how to engage the massive edge devices in large-scale computation also needs to be studied. Effective computing paradigms are significant for collaborating numerous edge devices through low bandwidth and making small devices useful in big-model computing.

\subsection{Low-Delay Inference}
There is always a high pursuit for real-time in wireless communication, and the information expiration caused by high latency also harms transmission accuracy. After introducing wBAIM into wireless systems, the inference delay of the model and the interaction delay between the model and the other wireless components, such as RF modules, will also become part of the total delay. How to make wBAIM perform fast inference and closely link RF components and AI models is important. In addition, it is also valuable to research how to utilize the intelligence of wBAIM to predict and compensate the state changes to offset the harm caused by delay.

\subsection{Working with Restricted Interaction}
Subject to the limited communication resources, the information interaction between users and BS in a short period, such as a coherent time, is limited. How to perform as efficient information transmission and utilization as possible under interaction constraints? This involves the following sub-techniques: establishing a matched information compression and reconstruction scheme at the transmitter and receiver, such as the encoder and decoder pairs in MIMO CSI feedback \cite{csinet}; designing the wBAIM so that it can perform feature extraction despite receiving only partial information as input; driving the wBAIM to output high-density and user-friendly information, which is the special case of semantic communication in transmitting physical layer information.

\subsection{Mitigating Interference, Noise and Errors}
Interference, noise, and error are constant challenges to wireless communications and will carry over to the wireless models. In practical applications, the signals received by wBAIMs will be noisy, the signals sent by wBAIMs will be biased during transmission, and the acquired state information will be inaccurate. This makes the research on wBAIMs need to concern several issues: how to improve the denoising ability of the model through effective structural designs; how to exercise the robustness of the model to cope with potential error perturbations through suitable training process; how to use the model's intelligence to estimate the interference among multiple users and adaptively select the anti-interference scheme.

\renewcommand{\arraystretch}{1.35}
\begin{table*}[]\small
\caption{\small Some research directions and potential solutions for wBAIMs}
\centering

\begin{tabular}{l|l|l}
\hline \hline
\makecell[c]{Process} & \makecell[c]{Research   directions}                 & \makecell[c]{Potential   solutions}                                  \\ \hline \hline
\multirow{5}{*}{Dataset creation}               & \multirow{2}{*}{Data collection}          & Privacy protection: quantum protocol, blockchain, homomorphic encryption                                                 \\ \cline{3-3} 
        &          & Fairness and diversity:  scheduling policy                                       \\ \cline{2-3} 
                                                & \multirow{3}{*}{Data   augmentation}      & Priori based methods:   complex-domain rotation, cropping, flipping \\ \cline{3-3} 
        &                                       & Simulation datasets based on wireless model            \\ \cline{3-3} 
        &                                       & Learning-based methods: GAN, VAE, diffuison model      \\ \hline
\multirow{4}{*}{Model   design}                 & \multirow{2}{*}{Model structure}          & Transformers                                                         \\ \cline{3-3} 
        &                                       & Physics-inspired methods: multi-dimensional RNN, Siren \\ \cline{2-3} 
        & \multirow{2}{*}{Objective   function} & Autoregressive learning                                  \\ \cline{3-3} 
        &                                       & Masked learning                                        \\ \hline
\multirow{4}{*}{\makecell[l]{Training on \\ wireless networks}} & \multirow{2}{*}{Federated learning}       &         Asynchronous and heterogeneous architecture                                                            \\ \cline{3-3} 
        &                                       &     Multi-task  structure                                                  \\ \cline{2-3} 
        & \multirow{2}{*}{Split learning} &    Over-the-air computation for split learning 
        \\ \cline{3-3} 
        &                                       &  Federated split learning                                                     \\ \hline
\multirow{3}{*}{Deployment}                     & \multirow{2}{*}{Low-cost and low-latency} & Parameter   quantization                                            \\ \cline{3-3} 
        &                                       & Sample-wise adaptive inference                         \\ \cline{2-3} 
        & Seamless on wireless system           & Integrating intelligent computation, radio, and networking   \\ \hline \hline
\end{tabular}
\label{table_research}
\end{table*}

\subsection{Multiple Access Service} 
Distinguishing multi-users with the multi-access strategy is a unique feature of wireless networks and is also needed in the functionality of wBAIMs. For existing BAIMs, the input and output are the information filtered out of the user address characteristics. For example, if different users ask the same question, then chatGPT receives the same token sequence as input and can also answer the same token sequence as output for different users. However, for the wBAIM, which works directly in wireless networks, its information processing and expression cannot ignore the differences between user addresses. For example, in the FDMA mode, the CSI of different users differ not only in the channel path component but also in the allocated subcarriers. In this case, when inputting CSI of multi-user into the wBAIM, the model should be able to additionally distinguish the characteristics difference between multiple frequency addresses; when the wBAIM outputs multi-user information, it should add the corresponding frequency address characteristics. For this unique challenge, there are two ideas worth investigating. The first is to make the model work independently of the physical domain used to distinguish multiple users and exclude the differentiation between user signals. Another one is to make the model accept the user’s address information embedding to directly provide differentiated service tailored to user’s address.

\section{Research Directions and Potential Solutions for wBAIMs}\label{section_researches}

The general deep learning solutions include the following procedures, i.e., dataset creation, model design, training, and deployment. How to build the anticipated wBAIM-based wireless architecture and how to solve the existing challenges will also be concretely implemented in these processes. This section discusses some promising research directions and potential technical approaches in these four processes, as summarized in Table \ref{table_research}.

\subsection{Building Massive, Informative, and Heterologous Wireless Datasets}

\subsubsection{Real Data Collection} 
The dataset serves as an informative source, imparting knowledge to the model and guiding the learning process. Therefore, to reduce the shift in the practical deployment and training, the training data should preferably be collected from practical systems. Further, protecting privacy and ensuring fairness are imperative when collecting data in practical systems. Quantum protocols, blockchain, homomorphic encryption, and scheduling policies are potential technological tools.

\subsubsection{Data Augmentation}
After data collection, augmenting wireless data to make limited data yield value equivalent to more data is also valuable. One way is to expand the legitimate data, such as complex-domain rotation, cropping, and flipping, based on a priori physical properties of electromagnetic signals. Another way is to generate simulated datasets based on known wireless models, such as channel-position pairs based on ray-tracing models. Moreover, it is also possible to first learn the distributional properties of collected data using generative models such as generative adversarial network (GAN) \cite{mimo_gan} and then expand the training data based on the learned models. 

\subsection{Wireless Characteristic Representation and Modelling}

\subsubsection{Model the Wireless Channel Structure and Properties}
The structure of the model determines its capability in feature capture and representation. The current BAIMs are generally based on transformers. The attention mechanism in the transformer is also valuable for learning wireless properties, especially for characterizing the correlation in spatial, temporal, and frequency domains. Also, physics-inspired wireless AI methods are noteworthy, such as multidimensional recurrent neural network (RNN) for learning the structure of high-dimensional channels \cite{2DSeq2Seq} and periodic activation function for fitting the phase of electromagnetic waves \cite{siren}.

\subsubsection{Derive the Objective Function for a Wireless Context}
The objective function significantly affects the expression form of the learned intelligence. There are two typical objective functions of BAIMs: autoregressive and masked learning. Autoregressive learning \cite{gpt3} is well suited to portray communication processes that occur continuously in time and thus has also been used for tasks such as channel prediction.  Masked learning \cite{bert}, which masks a portion of the input and drives the model to infer the masked information, is beneficial for enhancing supervised information in unlabeled data. 
Also, masks simulate partial input, which can enhance the learning ability in restricted interactions and the model's robustness. Alternatively, masks can be used for address information embedding to achieve the multiple access service.

\subsection{Centralized and Distributed Training over Wireless Networks}

\subsubsection{Federated Learning for Connecting Data Nodes}
Federated learning \cite{10024766} uses model or gradient swapping instead of data swapping, which is especially important for protecting data privacy, addressing data silos, and mobilizing the computational resources of edge devices. Also, asynchronous and hierarchical schemes are of particular significance due to the huge number of participating devices and data nodes in the training of wBAIM. Meanwhile, the federated multi-task structure is also significant in enhancing the versatility of the wBAIM.

\subsubsection{Split Learning for Combining Computing Devices}
Splitting big models to train on multiple devices will help mobilize the computing power of small computing devices in wireless systems. In particular, the framework that combines over-the-air computing with splitting learning \cite{kim2023bargaining} not only facilitates splitting learning in wireless systems but also naturally introduces wireless noise in training to exercise the robustness of the model. Besides, federated split learning, a framework for combining both data nodes and computing devices, is also meaningful for training the wBAIM over wireless networks.

\subsection{Low-Cost, Low-Latency and Seamless Deployment}

\subsubsection{Parameter Quantization}
One way to optimize the inference algorithm is to simplify the model, such as quantizing the parameters to reduce the model complexity. Although slightly harming the accuracy of the inference results, a proper quantization method will significantly reduce the inference latency as well as the storage and memory overhead.

\subsubsection{Sample-Wise Adaptive Inference}
Another way to reduce the inference cost and delay is to perform sample-wise adaptive inference. If the belief of the inference results at some layer/submodule is high, the forward computation can be discontinued. This allows the inference to be determined based on the difficulty of the sample, adaptively adjusting the amount of computation for each sample.

\subsubsection{Integrating intelligent computation, radio, and networking} 
The wBAIM needs to be coordinated with the existing wireless system, especially the necessary radio hardware and networking technologies, to fully unleash its application value. Establishing multi-level information links between radio components and intelligent computing units can sufficiently intelligentize the transmission and reduce the total inference latency. Besides, collaboration with advanced networking technologies, such as software-defined networking (SDN) \cite{sdn}, is significant for applying wBAIM to intelligent scheduling and massive access, as a flexible network architecture can facilitate the deployment of real-time customized solutions.

\section{Conclusion}\label{section7}
This paper provides an outlook on the BAIMs for 6G wireless networks and outlines the opportunities, challenges, and research directions. The meeting between the development of AI technology and the evolution of wireless systems is expected. Finally, we offer some evaluative development recommendations: 
\begin{itemize}[leftmargin=*]
\item \textbf{Recommendation 1:} The primary goal of research on the wBAIM is not simply to scale up the neural networks but to establish a unified multi-task and scenario intelligent model and deployment paradigm. 
\item \textbf{Recommendation 2:} Research should pay attention to the additional constraints brought by the specificity of wireless networks, making AI technologies act according to wireless circumstances.
\item \textbf{Recommendation 3:} The synergistic development of software and hardware is indispensable, and a network system that seamlessly couples computing and communication to support AI is also critical.
\end{itemize}

\ifCLASSOPTIONcaptionsoff
  \newpage
\fi

\bibliographystyle{IEEEbib}
\bibliography{ref}

\end{document}